\documentclass[draft]{agujournal2019}
\usepackage{url} 
\usepackage{lineno}
\usepackage[inline]{trackchanges} 
\usepackage{soul}
\draftfalse

\journalname{Geophysical Research Letters}

\begin{document}

%
%


\title{Physically Interpretable Emulation of a Moist Convecting Atmosphere with a Recurrent Neural Network}

%
%




\authors{Qiyu Song\affil{1}, Zhiming Kuang\affil{1,2}}


\affiliation{1}{Department of Earth and Planetary Sciences, Harvard University, Cambridge, MA, USA}
\affiliation{2}{John A. Paulson School of Engineering and Applied Sciences, Harvard University, Cambridge, MA, USA}




\correspondingauthor{Qiyu Song}{qsong@g.harvard.edu}



\begin{keypoints}
\item We present the first data-driven convection parameterization using a recurrent neural network that includes convection memory.
\item The model exhibits stable and realistic behavior in long-term emulations when coupled with two-dimensional gravity waves.
\item State-dependent linear responses to perturbations offer interpretable insights into the convectively coupled wave system.
\end{keypoints}

%
%

%
%


\begin{abstract}
Data-driven convective parameterization aims to accurately represent convective adjustments to large-scale forcings in a computationally economic manner. While previous studies have demonstrated success using various model architectures, challenges persist in developing physically interpretable models and assessing generalizability and confidence level. In this study, we develop a recurrent neural network to predict time series of temperature, moisture, and precipitation of a cumulus ensemble in response to large-scale forcings. The recurrent cell combines a linear component, pre-identified as a time-invariant state-space model within the linear limit of the problem, and a multilayer neural network for the nonlinear component. Trained on ensembles of limited-domain cloud-resolving model simulation data, the model exhibits stable and realistic performance in long-term emulations, both with prescribed large-scale forcings and when coupled with two-dimensional gravity waves. We further calculate linear responses to perturbations for the coupled emulation, revealing physically interpretable, state-dependent properties of the convectively coupled gravity wave system.
\end{abstract}

\section*{Plain Language Summary}
Understanding how convective systems interact with broader-scale atmospheric patterns is crucial to improving weather forecasts and climate models. In this study, we developed a new machine-learning approach to predict changes in temperature, moisture, and rainfall based on large-scale atmospheric conditions. The model explicitly includes memory from the past and passes it through time to enhance future predictions. After being trained on high-quality data from carefully designed simulations, the model successfully achieved stable and realistic long-term simulations. This new model not only provides accurate predictions but is also interpretable and can offer a better understanding of the underlying physical processes.

%
%

%


%
%
%
%

\section{Introduction}
Due to the small spatial scales of atmospheric convective processes compared to climate model resolutions, parameterizations are necessary for representing the interaction between moist convection and the large-scale flow. Current global climate models (GCMs) use physics-based convective parameterizations, which remain major sources of uncertainty in climate simulations and projections \cite{IPCC}.
High-resolution models such as cloud-resolving models (CRMs) offer improved accuracy by explicitly simulating clouds and convection at finer resolutions ($<$5 km), but are computationally expensive for long-term global simulations \cite{globalSAM, taylor2023simple, hohenegger2023icon}.
The Multiscale Modeling Framework (MMF) provides an alternative by embedding small-domain CRMs within GCM columns as a parameterization \cite{khairoutdinov2001cloud, randall2003breaking}.
The MMF has been shown to reduce model biases and better represent phenomena such as the convectively coupled waves and the Madden-Julian Oscillation \cite{khairoutdinov2008, benedict2009mjo, arnold2015, kooperman2016}, but still requires significantly more computational resources than GCMs.

To overcome the accuracy–cost trade-offs of convection modeling, machine learning (ML) has emerged as a powerful, data-driven alternative. By learning complex input–output mappings directly from data, ML-based schemes can capture the complicated physics of convection at high fidelity and reduced computational expense.
Previous studies have trained various model architectures with global CRM/MMF simulations \cite{gentine2018could, rasp2018deep, brenowitz2018prognostic, brenowitz2020interpreting, yuval2021use} or reanalysis datasets \cite{neuralgcm}. Typically, these models predict convective tendencies based on thermodynamic and large-scale forcing profiles in atmospheric columns, sometimes also incorporating sensible heat, latent heat, and radiation fluxes, only from the current or from the current and a few prior time steps.

An underlying assumption for many parameterizations, as noted in \citeA{kuang2024}, is that convective adjustment occurs rapidly relative to large-scale forcing, maintaining quasi-statistical equilibrium. However, this assumption breaks down when large-scale forcing varies on timescales comparable to convection, because finite response time introduces convection memory. Convection memory has been suggested to have a stronger damping effect on shorter-period signals \cite{emanuel1994, kuang2008b} and affect convection lifecycle and organization \cite{cohen2004,conv_mem2}. Therefore, it is valuable to properly represent convection memory in parameterizations.

Previous attempts to model memory effects have incorporated various prognostic variables, but there is no consensus on whether or how this should be implemented physically \cite{pan_randall_1998, conv_mem1}.
From a data-driven perspective, \citeA{han2020moist} developed a residual convolutional neural network using up to four prior time steps to capture memory effects.
As an initial step towards a more systematic approach, \citeA{kuang2024} identified a linear time-invariant state-space model from carefully designed simulations. The use of the state-space model, especially when the dimension of its state vector exceeds that of the output domain-mean thermodynamic profiles, can represent the convection memory effect. \citeA{kuang2024} further introduced a procedure to interpret the memory effect physically. Based on well-established theories and approaches targeting linear problems, \citeA{kuang2024} provides a reference point for data-driven convection modeling within the linear limits, assuming small external forcing and deviations from the radiative-convective equilibrium (RCE).

In this work, we develop a nonlinear model that allows the atmosphere to deviate substantially from RCE. We propose a recurrent neural network (RNN) architecture \cite{RNN} that explicitly incorporates memory, similar to the linear state-space model in \citeA{kuang2024}. Our recurrent cell combines linear and nonlinear components, inspired by residual neural networks \cite{resNN}.
The model is initialized with a pre-identified linear component as in \citeA{kuang2024}, and nonlinear parameters are trained using simulation data from an ensemble of limited-domain CRMs under random forcing, as well as simulations that include coupling with large-scale gravity waves.
The model predicts time series of thermodynamic profiles (temperature and moisture) and surface precipitation in response to input series of large-scale thermodynamic forcings.
It integrates all physical processes within the atmospheric column, including moist convection, idealized radiation, and boundary layer processes.
In online tests where the RNN is coupled to two-dimensional gravity waves, the model achieves stable long-term emulations and accurately captures the nonlinear behaviors of the convectively coupled waves in the CRM simulations. Linearizing around varying reference states along wave trajectories further reveals time-evolving responses to perturbations with state-dependent features, offering a pathway towards physical interpretability.

The rest of the manuscript is organized as follows. Section \ref{sec:method} describes the training data, including the CRM used to generate data and the experimental designs, the RNN model architecture, and the parameter learning process. Section \ref{sec:results} presents the offline performance, online test results, and physically interpretable linear approximations of the model. Finally, Section \ref{sec:discussion} provides summary and discussion.

\section{Methods} \label{sec:method}
\subsection{Data Generation}
The training and validation data are generated with the System for Atmospheric Modeling (SAM), version 6.11.7, as described by \citeA{SAM_model}.
SAM employs an anelastic dynamical core and uses liquid water static energy, total non-precipitating water, and precipitating water as prognostic thermodynamic variables. A simple Smagorinsky-type 1.5-order scheme is used to parameterize subgrid-scale turbulence.
Radiation is idealized following \citeA{idealized_rad}, with a constant cooling rate of $-1.5~\mathrm{K/day}$ in the troposphere and Newtonian relaxation to $200~\mathrm{K}$ for temperatures below $207.5~\mathrm{K}$, with a time scale of 5 days.
Surface fluxes of latent and sensible heat are modeled using a bulk aerodynamic formula with a constant 10-meter exchange coefficient of $1\times 10^{-3}$ and a fixed surface wind speed of $5~\mathrm{m/s}$. The sea surface temperature is fixed at $29^\circ$C across the domain.
A sponge layer is employed in the upper third of the domain to reduce gravity-wave reflection.
The model includes 28 vertical levels extending from the surface to approximately $32~\mathrm{km}$, with a horizontal domain covering 128$\times$128 km at 4-km resolution. These resolutions align with those of the CRM in the Superparameterized Community Atmosphere Model (SPCAM), an MMF climate model so that work described here could be extended in the future to SPCAM.

In each experiment, the model is first run without external forcing for 200 days to achieve RCE, followed by 50 days to collect RCE mean-state data. Two groups of simulations are then conducted with different external forcings on temperature and moisture:

A. Random Forcing Simulations:

We use 1000-day sequences generated by randomly phased multisines as prescribed forcing. Across 18 experiments, the forcing magnitude ranges from 4 to 50, where a unit magnitude corresponds to changes over 15 minutes with a standard deviation of 0.0066 K in temperature or 0.066\% in relative humidity (RH). Scaled to daily values, the maximum magnitude represents approximately 0.6 K or 6\% RH change. Forcings of different layers and of temperature and humidity are independent of each other. Each simulation averages data from 256 ensemble members to enhance the signal-to-noise ratio, with all members sharing the same input forcing. Additionally, 3 experiments using an ensemble size of 1024 and forced by 5 repeating 200-day sequences with a magnitude of 4 are conducted to identify a linear model following \citeA{kuang2024} for initializing the RNN.

B. Coupled-wave Forcing Simulations:

We conducted 30 additional experiments in which the CRM domain is coupled to a two-dimensional (2D) gravity wave with a horizontal wavenumber $k$ ranging from 6 to 20, with a unit of $2\pi\times(40000~\mathrm{km})^{-1}$. Each experiment generates 150 days of data, averaged across 16 ensemble members. The forcing input $\mathbf{i}$ is calculated as advection from the vertical motion $w$ with a damping term, defined by:
\begin{equation}
    \mathbf{i}=\left[-w\left(\partial_z T+\frac{g}{c_p}\right),\quad -w(\partial_z q)\right]^T \Delta t - \epsilon \left[T',~q'\right]^T \Delta t
, \label{eq:i}
\end{equation}
where the evolution of $w$ follows:
\begin{equation}
    (\partial_t+\epsilon)\partial_{zz}(\bar\rho w) = -k^2 \bar\rho g \frac{T_v'}{\overline{T}_v}. \label{eq:w}
\end{equation}
Here, $\bar\rho$ represents the mean-state density, $\overline{T}_v$ is the mean-state virtual temperature, and $T$ and $q$ are the temperature and specific humidity profiles. The term $(\cdot)'$ denotes anomalies from the mean state, and $\epsilon$ is the damping coefficient. Using the linear wave equation as the convectively coupled waves saturate nonlinearly allows us, albeit artificially, to focus on nonlinearity within the behaviors of the cumulus ensemble. We further multiply the advection term by $1+r$, where $r$ is a random series drawn from the standard normal distribution and is independent for each channel. This reduces artificial correlation across channels and also generates training data points that produce a halo around the trajectory in the phase space. Such a halo allows the RNN training to include deviations from the unperturbed trajectory so that the RNN can better avoid the issue of extrapolation when it is run in the online mode. Forcing inputs and vertical velocity are updated every 15 minutes. The coupled equation for $w$ evolution is derived from perturbation equations of momentum, continuity, and hydrostatic balance, as described in \citeA{kuang2008a}.


\subsection{Model Architecture}
\begin{figure}
    \centering
    \includegraphics[width=1\textwidth]{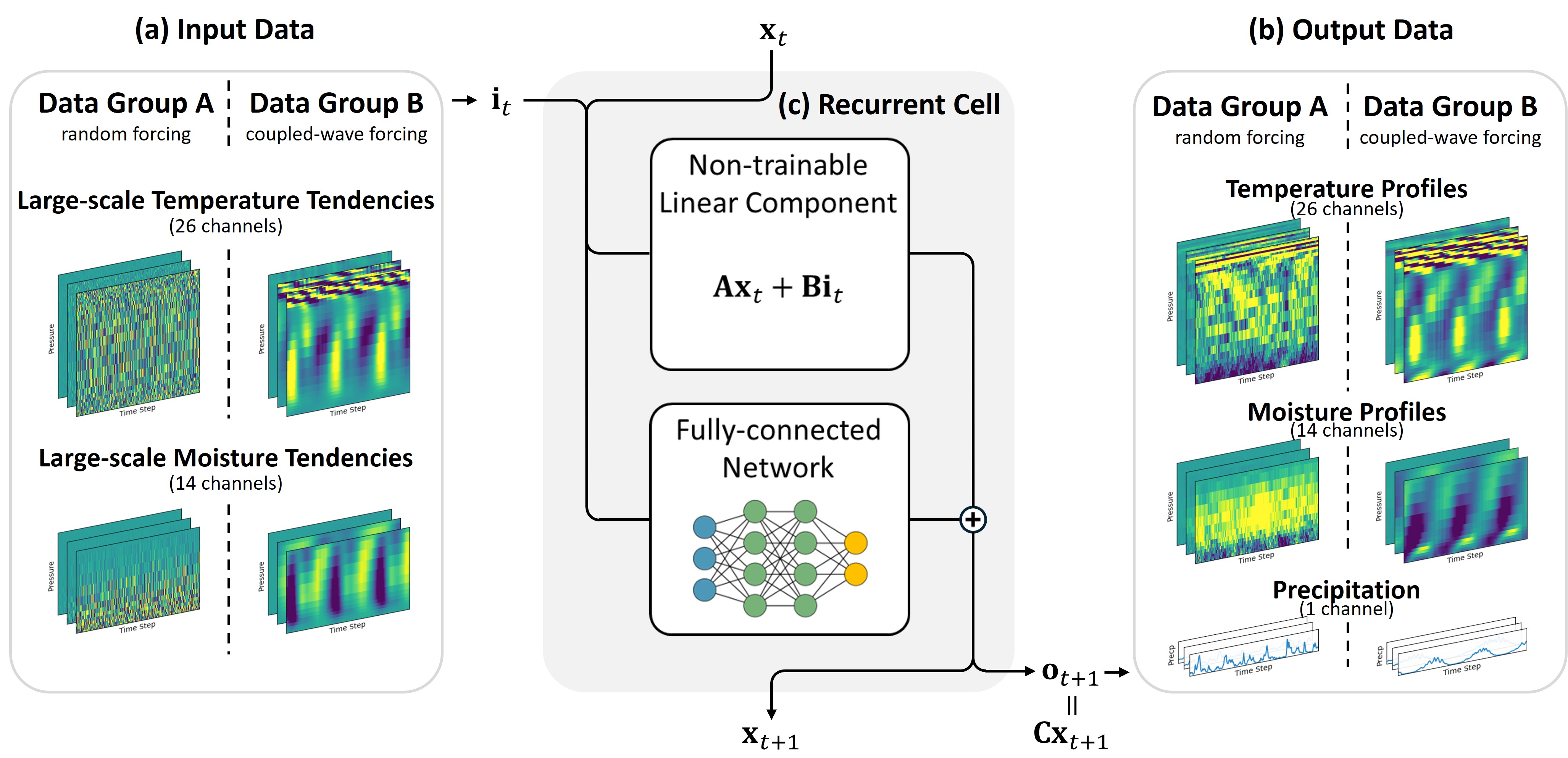}
    \caption{\textbf{Model Schematic.} \textbf{(a)} Input data includes 40 channels: 26 layers of large-scale temperature foricng and 14 layers of moisture forcing. Data comes from two groups of simulations with A) random forcing and B) coupled-wave forcing. \textbf{(b)} Output data includes 41 channels: 26 vertical layers of temperature anomalies, 14 vertical layers of moisture anomalies, and 1 channel for anomalous logarithmic precipitation.  \textbf{(c)} The recurrent cell: The state and input vectors are combined and passed through both a linear mapping and a 2-layer fully connected neural network, with the results summed to form the new state vector.  The output vector is diagnosed from the state vector using a linear mapping. Matrices $\mathbf{A}$, $\mathbf{B}$, and $\mathbf{C}$ are from pre-identified linear model and are non-trainable.}
    \label{fig:model_structure}
\end{figure}

We use a RNN architecture with a non-trivial recurrent cell in this work (Figure \ref{fig:model_structure}). The model's input is the large-scale forcing tendencies multiplied by the 15-minute model time step. The output includes temperature and moisture profiles after convective adjustment, along with an additional channel for surface precipitation. We consider 26 vertical layers for temperature and the bottom 14 layers for moisture, as specific humidity above 350 hPa has minimal effects with the idealized radiation used here. Temperature and moisture are weighted based on their individual contributions to energetic variability, as described in \citeA{kuang2024}. For surface precipitation, we apply a logarithmic transformation (with a $10^{-5}~\mathrm{mm/day}$ offset to avoid log-zero issue) and calculate its anomaly relative to the logarithm of mean-state precipitation in the RCE.

The core of the RNN model is the recurrent cell, which iteratively processes the hidden states from $\mathbf{x}_t$ to $\mathbf{x}_{t+1}$ as 64-dimensional vectors following the evolution equation:
\begin{equation}
    \mathbf{x}_{t+1} = \mathbf{A}\mathbf{x}_t+\mathbf{B}\mathbf{i}_t+\mathrm{NN}(\mathbf{x}_t,~\mathbf{i}_t), \label{eq:x}
\end{equation}
combining a linear mapping and a neural network (denoted as NN) with 2 hidden layers of width 312 and rectified linear unit (ReLU) activation function for all nodes. The output $\mathbf{o}_t$, comprising anomalies for temperature ($T'_t$), moisture ($q'_t$), and logarithmic precipitation ($P'_t$), is then diagnosed as
\begin{equation}
    \mathbf{o}_t \equiv \left[T'_t,\quad q'_t, \quad P'_t\right]^T = \mathbf{C}\mathbf{x}_{t}. \label{eq:o}
\end{equation}
The matrices $\mathbf{A}$, $\mathbf{B}$, and $\mathbf{C}$ are from the pre-identified linear model based on experiments with the smallest random forcing. This identification technique is well-established \cite<e.g.,>{Ljung1999} and has been applied to data-driven convection modeling by \citeA{kuang2024}. Only the parameters in the neural network are trainable, which are initialized as small random values drawn from a zero-mean Gaussian distribution with a standard deviation of $10^{-3}$.

There is a mathematically equivalent way that combines the linear mapping and the neural network as a single neural network (see supporting information). However, using a separate representation simplifies initialization and helps maintain behavior consistent with the pre-identified linear model.

\subsection{Parameter Learning}


During training, the model is first trained on experiments with random forcing. Subsequently, to improve performance in the case of the convectively coupled waves, data from experiments with coupled-wave forcing is added to the training. The training samples consist of 2-day slices extracted from the sequential data, with only the first 80\% samples from each experiment used for training. The loss function is the mean square error, scaled by the target output variance in each channel of each experiment. Details of the training process are provided in the supporting information. The whole training process takes approximately 50 hours on four A100 GPUs.

After training, we test the model on the entire dataset, covering the full sequences from all experiments. This comprehensive evaluation assesses the model's ability to predict steps not included in the training samples and tests its stability for long-term predictions.

\section{Results}\label{sec:results}
\subsection{Offline Performance}
\begin{figure}
    \centering
    \includegraphics[width=1\linewidth]{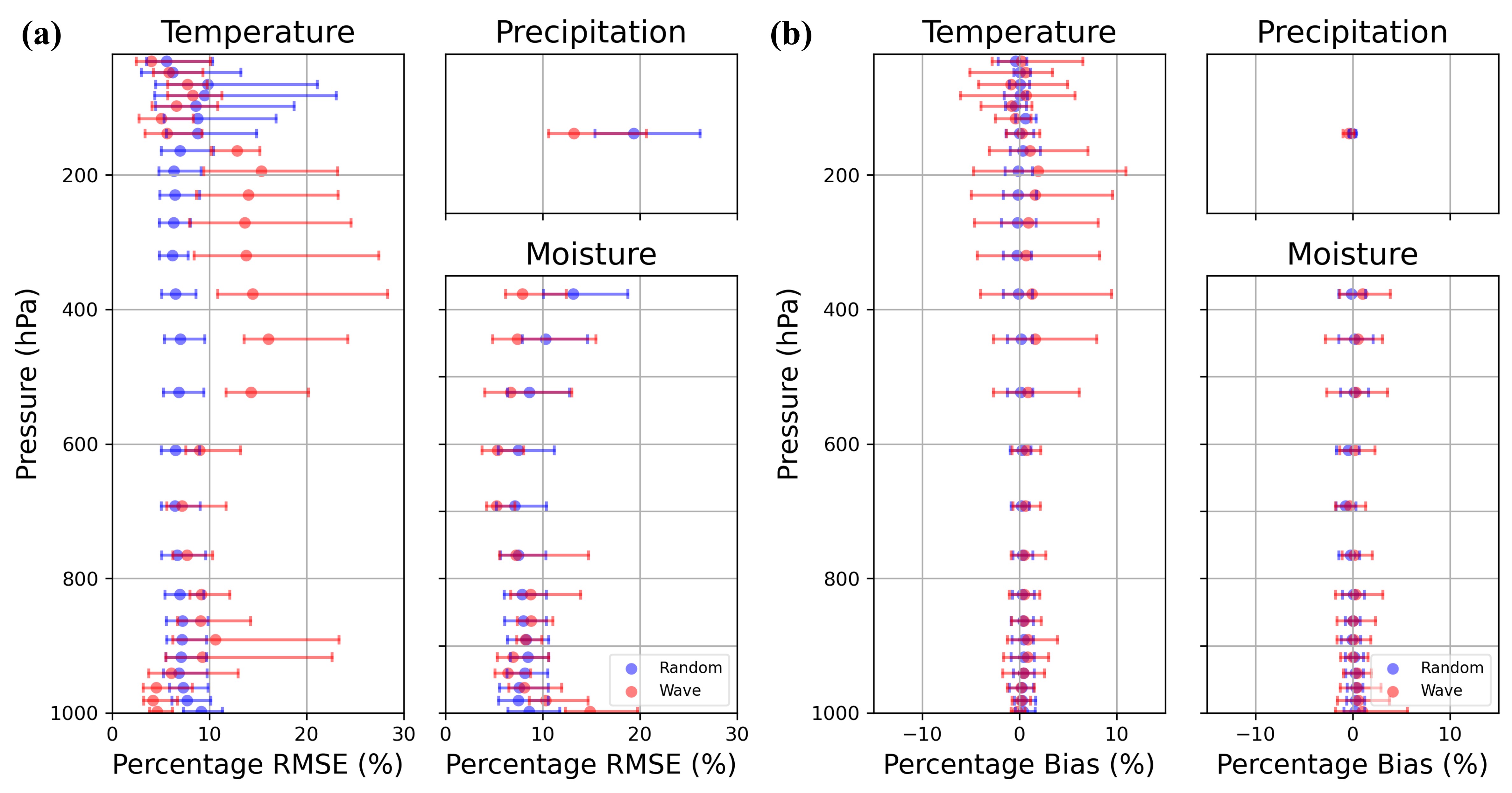}
    \caption{\textbf{Offline test performance.} \textbf{(a)} Percentage RMSE for predictions of temperature anomaly profiles, specific humidity anomaly profiles, and logarithmic precipitation anomaly in randomly-forced experiments (blue) and coupled-wave experiments (red).Horizontal bars at each level show the range across all experiments, with dots indicating the group average. \textbf{(b)} Same as (a), but for percentage bias of the mean state.}
    \label{fig:offline_test}
\end{figure}
Figure \ref{fig:offline_test} shows the offline performance of the learned model evaluated using two metrics: percentage root mean square error (RMSE) and percentage bias. The metrics are defined as:
\begin{eqnarray}
  \mathrm{Percentage~RMSE} & = & \sqrt{\frac{\sum_{t=1}^{N}\left(\mathrm{Prediction}(t) - \mathrm{Target}(t)\right)^2}{N\cdot\mathrm{Var}(\mathrm{Target})}}\times 100\%, \\
  \mathrm{Percentage~Bias} & = & \frac{\sum_{t=1}^{N}\left(\mathrm{Prediction}(t) - \mathrm{Target}(t)\right)}{N\cdot\mathrm{Std}(\mathrm{Target})}\times 100\%.
\end{eqnarray}
These metrics are calculated for each output channel in each experiment over the entire 1000-day or 150-day time series. In experiments with random forcing, the mean RMSE is generally below 10\%, except for moisture in the upper troposphere (15\%), and precipitation (20\%). These higher RMSE values are due to the small amount of moisture in the upper troposphere and the high variability in precipitation under random forcing conditions. Performance is comparable for experiments with coupled-wave forcing, where the mean RMSE is higher for temperature in the mid-to-upper troposphere and near-surface moisture (around 15\%), but lower for near-surface temperature, stratospheric temperature, mid-level moisture, and precipitation. This can be partially attributed to the heterogeneous variability across different channels and difference in vertical structure compared to random forcing experiments.

The mean bias remains mostly within $-2$\% to 2\% for both experiment groups, indicating that the model effectively captures changes in the time averaged states: generally, under higher-magnitude random forcing, stronger fluctuations in convection lead to warmer and moister time averaged states in the free troposphere due to asymmetries in the active and suppressed convective regimes. This performance extends that of the linear model in \citeA{kuang2024}, which is unable to capture the time averaged state changes.

\subsection{Online Performance}
\begin{figure}
    \centering
    \includegraphics[width=1\linewidth]{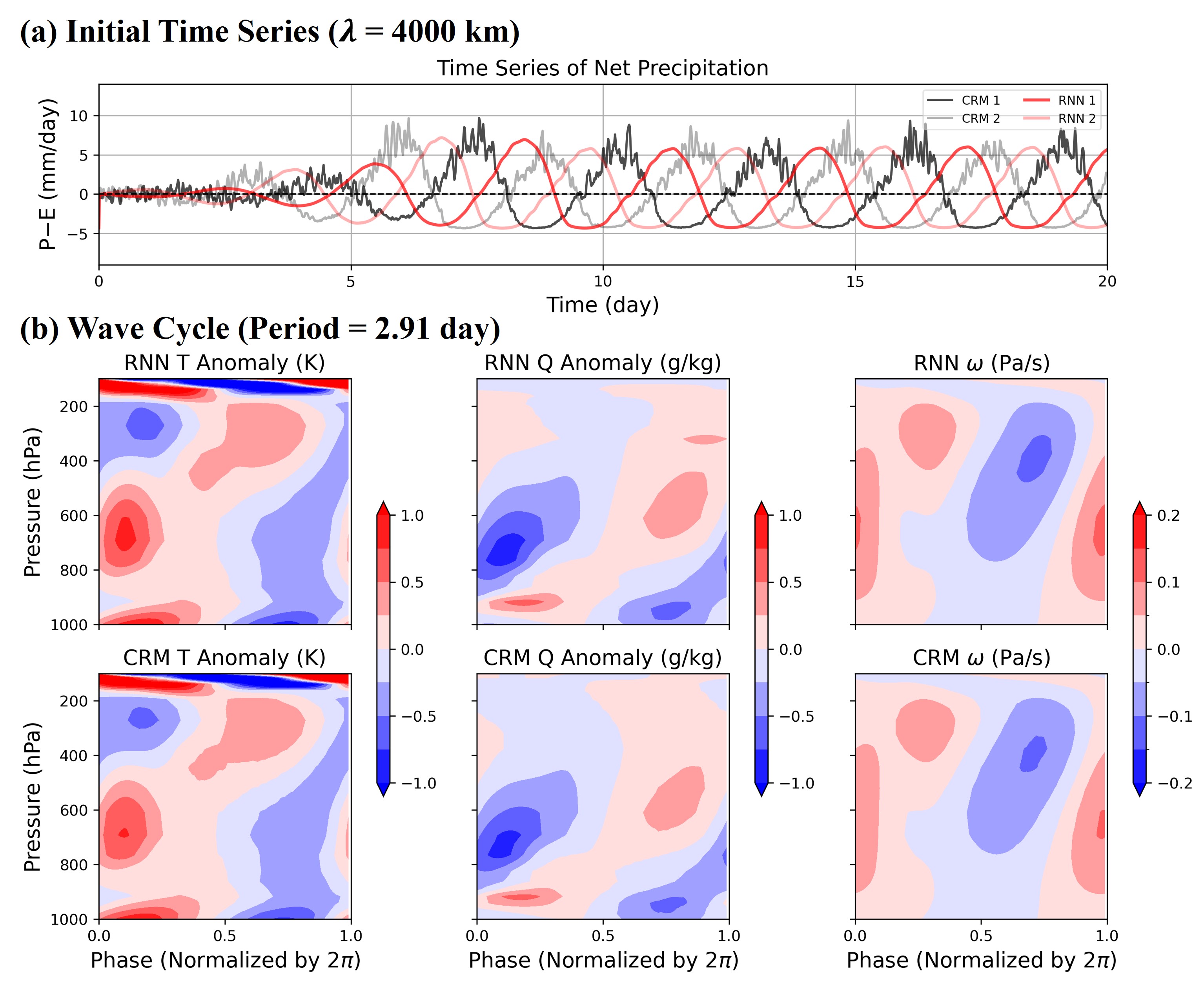}
    \caption{\textbf{Results for online coupled 4000-km wave experiment.} \textbf{(a)} Time series for net precipitation (precipitation minus evaporation) over the first 20 days for 2 CRM simulations (black and gray) and 2 RNN emulations (red). Phases do not match because of random initial states. \textbf{(b)} Extracted composite cycle for the 2.91-day period waves for both RNN emulation (top row) and CRM simulation (bottom row), showing temperature anomaly, specific humidity anomaly, and pressure velocity profiles. Phases in \textbf{(b)} are temporally aligned for comparison between the two models.}
    \label{fig:wn10}
\end{figure}
To evaluate the model's online performance when coupled with 2D gravity waves, we solve Eqn. (\ref{eq:i}) to (\ref{eq:o}) in discrete time, initializing the state vector with random values and using a damping time scale of 2 days ($\epsilon=0.5~\mathrm{day^{-1}}$). The coupled experiments run stably for at least 1000 days, closely repeating a wave cycle.

Figure \ref{fig:wn10} compares online emulations with CRM simulations for the 4000-km wave. Our model accurately captures the growth and saturation of the wave amplitude. The range of net precipitation (precipitation minus evaporation) closely matches CRM results, exhibiting asymmetric behaviors: (a) a relatively flatter bottom compared to the top of the oscillation, and (b) a longer increasing phase of net precipitation compared to the decreasing phase. After adjusting for phase shifts caused by random initializations, wave cycles from the RNN and CRM align well, resembling the convection life cycle described in previous studies \cite<e.g.,>{mapes2000}. The shown cycles begin at the strongest large-scale subsidence, with precipitation peaking after 66\% of the cycle. Similar results for 2000-km wave are included in supporting information.

\subsection{Linear Impulse Responses}
To interpret the behaviors of the RNN, we analyze its responses to small impulses in input forcing at different phases of wave oscillation. Starting from a given state, an impulse $\delta \mathbf{i}_t$ is applied as a single-layer perturbation to temperature or moisture input during one step ($\Delta t=15~\mathrm{min}$), in addition to the input forcing of the unperturbed simulation. The responses are calculated as differences between the perturbed and unperturbed outputs after $n$ steps, excluding the impulse itself. This can be computed by linearizing Eqn. (\ref{eq:x}) in combination with Eqn.(\ref{eq:o}):
\begin{equation}
    \delta \mathbf{o}_{t+n} - \delta \mathbf{i}_t = \left(\mathbf{C} \left(\mathbf{A}+\left.\frac{\partial \mathrm{NN}}{\partial \mathbf{x}}\right|_{\mathbf{x}_t, \mathbf{i}_t}\right)^{n-1} \left(\mathbf{B}+\left.\frac{\partial \mathrm{NN}}{\partial \mathbf{i}}\right|_{\mathbf{x}_t, \mathbf{i}_t}\right)-\mathbf{I} \right) \delta \mathbf{i}_t.
\end{equation}
where $\mathbf{I}$ is the identity matrix. This equation is valid for small $\delta \mathbf{i}_t$ and small $n$, allowing local linearization around a given state.

\begin{figure}
    \centering
    \includegraphics[width=1\linewidth]{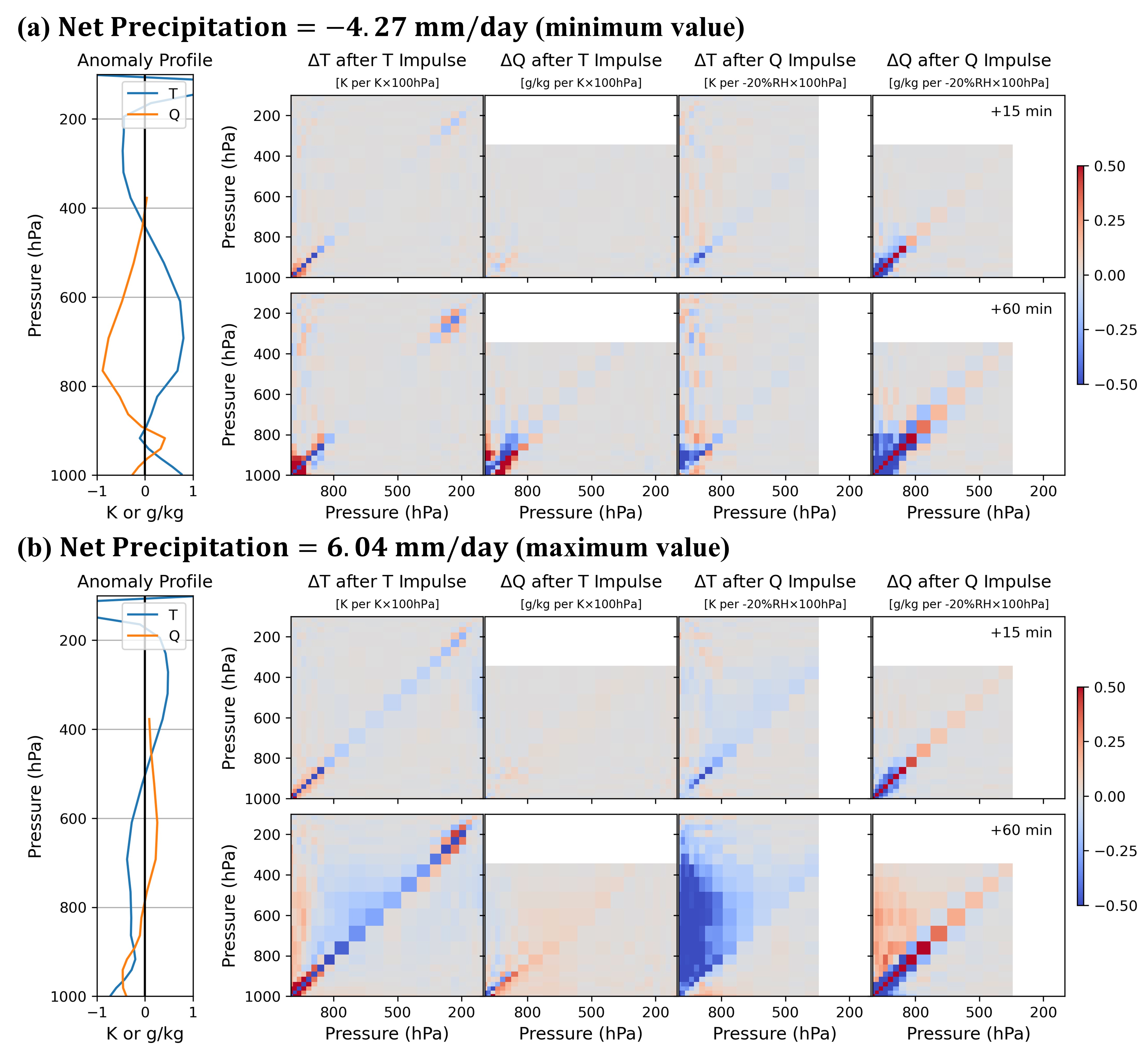}
    \caption{Linear impulse response functions of the states of (a) minimum and (b) maximum net precipitation along the trajectory of the simulated coupled waves with 4000-km wavelength. In both (a) and (b), the left panel shows the anomaly profiles for temperature and specific humidity, and the right panels show responses after 15 minutes (1st row) and 60 minutes (2nd row). The horizontal axis is the impulse pressure level, and the vertical axis is the response pressure level. Results are normalized by the mass of the perturbed layer to a 100-hPa layer mass equivalent.}
    \label{fig:response_4000km}
\end{figure}

Figure \ref{fig:response_4000km} illustrates linear impulse responses at the phases of minimum and maximum net precipitation in the 4000-km wave experiment. A key observation is the state-dependence of these responses. Under low-precipitation conditions, responses are largely confined to the lower troposphere, corresponding to a warm and dry mid-troposphere that suppresses deep convection. Conversely, under high-precipitation conditions, deep convection signals propagate throughout the troposphere. Linearization can be done around any states along the wave trajectory, with responses varying between these two regimes. Analyses for other wavelengths yield similar results.


\section{Discussion and Summary} \label{sec:discussion}
In this work, we used high-resolution, small-domain ensemble CRM simulation data to train an accurate and stable ML emulator. To our knowledge, this is the first application of an RNN model in the temporal dimension to capture convection memory for data-driven convection parameterization. Furthermore, we showed that the ML emulator is physically interpretable through trajectory piece-wise linear approximations.

Building on the concepts introduced by \citeA{kuang2024}, our model incorporates a state vector with a higher dimension than the output vector, enabling it to capture more information, including convection memory. The idea is extended into the nonlinear regime, making it applicable to states that deviate substantially from RCE. Compared to the model in \citeA{han2020moist}, which uses up to four prior time steps as input for memory effects, our model autonomously learns what information to retain and propagate across time steps. While the linear precursor of this model could be reformulated as a finite-step model \cite{kuang2024}, such an analysis for the current nonlinear model remains a task for future studies. This prognostic form with additional degrees of freedom somewhat aligns with the concept of reservoir computing \cite{reservoir1, reservoir2}. Unlike their model, which optimizes a diagnostic matrix with a fixed recurrent cell built from random matrices, our model optimizes state evolution directly. Additionally, our RNN is initialized using a pre-identified linear model, offering a robust and simplified starting point for optimizing its nonlinear components. The decomposition into linear and nonlinear components is conceptually similar to \citeA{brenowitz2018prognostic}, but our use of a pre-identified linear component reduces training complexity.

Regarding training data, many studies use more realistic simulations or reanalysis products, which only sample a subset of the hyperdimensional space of large-scale forcing. Approaches such as including warmer climate data \cite{rasp2018deep} or employing transfer learning \cite{sun2023data} improve out-of-distribution generalization but fail to fully address departures from the training manifold. This issue is particularly pronounced in dynamically coupled scenarios \cite{brenowitz2020interpreting}, where model may provide unstable or unbounded predictions, leading to instability. In this work, we train on randomly forced simulations, broadly sampling the entire hyperdimensional space. This strategy enhances generalizability and stability by incorporating cases that may be rare in realistic simulations but could arise due to compounded errors when coupling a neural network with large-scale dynamics. Adding data from coupled-wave simulations during the second training stage further improves the model’s performance in the targeted scenarios while preserving generalizability. Furthermore, introducing noise into the wave forcing prevents the model from learning noncausal relationships from spurious correlation among input channels.

In terms of physical interpretation, our model interrogate convective responses to perturbations in large-scale forcing around varying reference states. Unlike previous linear response functions \cite{kuang2010,kuang2012,kuang2018}, which examine convective tendencies under steady-state perturbations, our results reveal state-dependent, time-evolving dynamics. Qualitatively, while the RCE linear response functions align with the responses observed 4 steps after impulse in high-precipitation states, they differ significantly in low-precipitation states. Compared to previous linear response analyses for NN models \cite<e.g.,>{brenowitz2020interpreting}, our responses are less noisy. This may result from the use of a well-identified linear model as a foundational component and the incorporation of randomness in input forcing data during training. However, whether this continues to hold when a greater range of conditions is included, as in \citeA{brenowitz2020interpreting}, remains to be seen.

We acknowledge several areas for improvement in this work. While our model extends beyond the linear limit, it is still applicable only within a certain range around RCE. This corresponds to an instantaneous precipitation rate no more than $20-25~\mathrm{mm/day}$, or approximately $4-5$ times the average precipitation rate of $4.3~\mathrm{mm/day}$. Also, the 4-km resolution used in the CRM is relatively coarse compared to developing MMF models \cite<e.g.,>{sp-e3sm}, and several studies suggest that a much finer resolution would be needed to reach numerical convergence \cite{ramosvalle2023grid, hu2024refined}. Additionally, our model is deterministic. Although variability can be introduced through different initial states, this effect dissipates after a few days, or only causes a phase shift in coupled-wave emulation. A probabilistic approach, predicting distributions of variables instead of single deterministic values, could be beneficial. The hyperparameters used in training—such as network width and depth, sample extraction, batch size, and optimizer settings—were only loosely tested and may not be optimal. Systematic exploration of these configurations could further improve performance. Nonetheless, our results demonstrate the feasibility of modeling moist convection with memory effects using an RNN and merits further exploration for future development in ML-based convection models.

Future work may progress along two paths. To enhance the model's applicability in climate models, additional training is necessary. Currently, the model is limited to  tropical oceans with idealized radiation and surface fluxes over a single sea surface temperature. While this setup is already valuable for studying the nonlinear saturation of convectively coupled waves, expanding training to include more realistic physics and a broader range of atmospheric states will be essential for wider applicability. With such an extension, the ML model could replace the CRM in MMF models like SPCAM. Moreover, although the linearized representations offer an interpretable form of the system, which could be used for revealing various properties of the convectively coupled atmosphere as in previous studies \cite{kuang2010,kuang2012,kuang2018}, this is not yet done. This could involve reducing model order \cite{kuang2018} and reformulating into finite-step models like vector autoregressive models with exogenous inputs \cite{kuang2024}. These approaches represent valuable directions for future research.

\section*{Open Research Section}
The atmospheric model used to generate data is modified from the System for Atmospheric Modeling \cite{SAM_model} version 6.11.7, with the original version available at \url{http://rossby.msrc.sunysb.edu/SAM.html}. We use MATLAB version R2022b with the System Identification Toolbox to identify the linear model. We use Python version 3.10 with PyTorch version 2.0.1 to train the RNN model and perform analyses in this manuscript. The code for data generation and model training and testing, as well as all data needed to reproduce the training and analyses, are available at \url{https://doi.org/10.5281/zenodo.14648204} \cite{qiyu_song_2025_14648204}.

\acknowledgments
This work was supported by NASA Grant 80NSSC22K1837ZK and NSF Grant 1743753. The Harvard Cannon cluster provided the computing resources for this work. Q. S. thanks Zeyuan Hu for helpful discussion and comments.

%
%

\bibliography{references.bib}

%
%
%
%
%

\end{document}